\documentclass[apj, twocolumn]{openjournal}

\newcommand{\LCDM}{{$\Lambda$CDM}}
\newcommand{\e}[1]{\times 10^{#1}}
\newcommand{\msun}{\rm{M_\odot}}
\newcommand{\orcidauthor}[3]{\author{\href{http://orcid.org/#1}{#2$^{#3}$}}}

\usepackage{graphicx}	
\usepackage{xcolor}
\usepackage{amsmath}	
\usepackage{multirow}
\usepackage{enumitem}
\usepackage[breaklinks,colorlinks,citecolor=blue,urlcolor=blue]{hyperref}

\shorttitle{BCG Offsets in Cold Dark Matter}
\shortauthors{Roche et al.}

\begin{document}

\title{\vspace{-0.5cm}Brightest Cluster Galaxy Offsets in Cold Dark Matter\vspace{-1.5cm}}


\orcidauthor{0000-0002-3400-6991}{Cian Roche}{1,*}
\orcidauthor{0000-0001-5226-8349}{Michael McDonald}{1}
\orcidauthor{0000-0002-1327-1921}{Josh Borrow}{2,3}
\orcidauthor{0000-0001-8593-7692}{Mark Vogelsberger}{1,2}
\orcidauthor{0000-0002-6196-823X}{Xuejian Shen}{1}
\orcidauthor{0000-0001-5976-4599}{Volker Springel}{4}
\orcidauthor{0000-0001-6950-1629}{Lars Hernquist}{5}
\orcidauthor{0000-0003-3308-2420}{Ruediger Pakmor}{4}
\orcidauthor{0000-0002-0974-5266}{Sownak Bose}{6}
\orcidauthor{0000-0001-6092-2187}{Rahul Kannan}{7}

\affiliation{$^{1}$Department of Physics and MIT Kavli Institute for Astrophysics and Space Research, \\
77 Massachusetts Avenue, Cambridge, MA 02139, USA}
\affiliation{$^{2}$The NSF AI Institute for Artificial Intelligence and Fundamental Interactions, \\
77 Massachusetts Avenue, Cambridge, MA 02139, USA}
\affiliation{$^{3}$Department of Physics and Astronomy, University of Pennsylvania, 209 South 33rd Street, Philadelphia, PA, USA 19104}
\affiliation{$^{4}$Harvard-Smithsonian Center for Astrophysics, 60 Garden Street, Cambridge, MA 02138, USA}
\affiliation{$^{5}$Max-Planck-Institut für Astrophysik, Karl-Schwarzschild-Str. 1, D-85748, Garching, Germany}
\affiliation{$^{6}$Institute for Computational Cosmology, Department of Physics, Durham University, South Road, Durham DH1 3LE, UK}
\affiliation{$^{7}$Department of Physics and Astronomy, York University, 4700 Keele Street, Toronto, ON M3J 1P3, Canada}

\thanks{$^*$E-mail: \href{mailto:roche@mit.edu}{roche@mit.edu}}

\begin{abstract}
The distribution of offsets between the brightest cluster galaxies of galaxy clusters and the centroid of their dark matter distributions is a promising probe of the underlying dark matter physics. 
In particular, since this distribution is sensitive to the shape of the potential in galaxy cluster cores, it constitutes a test of dark matter self-interaction on the largest mass scales in the universe. 
We examine these offsets in three suites of modern cosmological simulations; IllustrisTNG, MillenniumTNG and BAHAMAS. For clusters above $10^{14}\,\msun$, we examine the dependence of the offset distribution on gravitational softening length, the method used to identify centroids, redshift, mass, baryonic physics, and establish the stability of our results with respect to various nuisance parameter choices. 
We find that offsets are overwhelmingly measured to be smaller than the minimum converged length scale in each simulation, with a median offset of $\sim1\,\rm{kpc}$ in the highest resolution simulation considered, TNG300-1, which uses a gravitational softening length of $1.48\,\rm{kpc}$. 
We also find that centroids identified via source extraction on smoothed dark matter and stellar particle data are consistent with the potential minimum, but that observationally relevant methods sensitive to cluster strong gravitational lensing scales, or those using gas as a tracer for the potential 
can overestimate offsets by factors of $\sim10$ and $\sim30$, respectively. This has the potential to reduce tensions with existing offset measurements which have served as evidence for a nonzero dark matter self-interaction cross section.
\end{abstract}

\keywords{\href{http://astrothesaurus.org/uat/265}{Cold dark matter(265)}; \href{http://astrothesaurus.org/uat/584}{Galaxy clusters(584)}; \href{http://astrothesaurus.org/uat/767}{Hydrodynamical simulations(767)}; \href{http://astrothesaurus.org/uat/1643}{Strong gravitational lensing(1643)}}

\section{Introduction}
The first evidence for the existence of a dominant and invisible matter component in the universe came from observations of the Coma galaxy cluster in 1933 \citep{zwicky:1933:ComaDM}, for which the kinematics of the galaxies within suggested an average density 400 times the inferred density from visible matter. The evidence for this dark matter (DM) has since accumulated to include for example the rotation curves of galaxies \citep{rubin:1970, ou:2023}, their escape velocities \citep{necib:2022}, the cosmic microwave background (CMB; \citealt{planck:2020}), and gravitational lens modelling \citep{clowe:2006, sharon:2020, Lin:2022, Natarajan:2024}. The particle nature of DM remains one of the primary questions of modern astrophysics, but a cold, collisionless DM in tandem with a cosmological constant $\Lambda$ constitutes the current standard cosmological model \citep{weinberg:2013, vogelsberger:2020:review}. 
The $\Lambda$ cold dark matter (hereafter \LCDM{}) cosmological paradigm has been remarkably successful in explaining the large scale structure of the universe via the Lyman-alpha forest \citep{evslin:2017, addison:2018, cuceu:2019}, the matter power spectrum and galaxy clustering \citep{reid:2010, eisenstein:2012:sdss}, the accelerated expansion of the universe \citep{riess:1998, perlmutter:1999} and observations of the cosmic microwave background anisotropies and polarizations \citep{page:2003}. 

Despite its successes, the \LCDM{} model has been subject to increasing tensions \citep{bull:2016, perivolaropolous:2022}, many of which relate to small scales such as the distribution, structure and diversity of the low-mass DM halos of dwarf galaxies (see \citealt{sales:2022} for a comprehensive review). 
Certain problems such as the ``core-cusp'' problem, in which CDM simulations predict greater dark matter density in the centers of dwarf galaxies than inferred from observations \citep{flores:1994, moore:1994}, can be resolved with either baryonic feedback \citep{navarro:1996:cores, weinberg:2015} or with the introduction of a nonzero dark matter self-interaction cross section \citep{spergel:2000, dave:2001,vogelsberger:2016}. 

The properties of DM halos and their baryons in the self-interacting dark matter (SIDM) paradigm have been well-studied in numerical simulations \citep[e.g.][]{vogelsberger:2012:sidm, rocha:2013, Zavala2013,Elbert2015,Robles2017, elbert:2018:sidmbaryonflattening,brinckmann:2018,zuhone:2019:gas,despali:2019,Fitts2019,robertson2019:bahamas,vogelsberger:2019,Shen2022,nadler:2023} 
but are generally described by a flattening of the potential in the cores of halos, particularly in DM-dominated systems like dwarf galaxies. For this reason most efforts in measuring the shape of density profiles as probes of the DM self-interaction cross section have focused on small scales, but significant opportunity also exists on large mass scales in the cores of galaxy clusters. 

\begin{figure*}
    \centering
    \includegraphics[width=\linewidth]{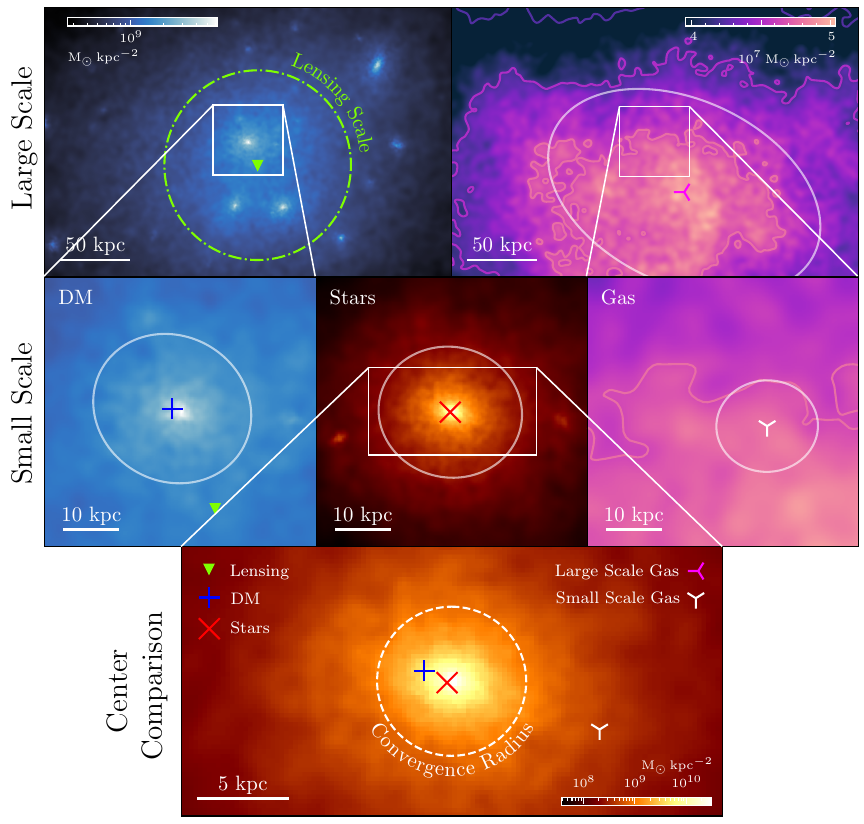}
    \caption{Demonstration of projected dark matter, gas and stellar density distributions in IllustrisTNG300-1 on scales relevant both to observations and to measuring faithful offsets between centroid measures of each distribution in simulations. All panels are centered on the group position (location of most bound particle) of group 20, which has a mass of $M_{200,\rm{mean}}=6.4\e{14}\,\msun$ and is shown at redshift zero. 
    \emph{(Upper left)} projected dark matter mass density on large scales in a cluster core, in addition to an isophote of the projected density map and its associated center (green). The selected isophote is matched in area to a circle of radius equal to 70kpc, which is the typical size of the Einstein radius for observations of a cluster on this mass scale. \emph{(Upper right)} the large-scale gas which has been used as a tracer of the potential, and its centroid as determined by SourceExtractor. The solid white ellipses in this panel and the middle row represent a scaled version of the source's elliptical shape, also determined by SourceExtractor.  \emph{(Middle row)} the small-scale dark matter, stellar and gas projected mass densities, and their centroids as determined by SourceExtractor. \emph{(Bottom)} comparison of the various centroid identification methods and their relationship to the convergence radius, given by $\sim2.8\,\epsilon_{\rm{DM,\star}}$ where $\epsilon_{\rm{DM,\star}}$ is the gravitational softening length of dark matter and star particles in the simulation (see text). The circle with radius equal to the convergence radius is centered on the most bound particle, ie. the center of the panel. The lensing scale center and large scale gas center are greatly offset from this position and are not visible on this panel.}
    \label{fig:demogrid}
\end{figure*}

Galaxy clusters form at the densest nodes in the cosmic web, with the highest dark matter densities in their cores. These cores constitute ideal testing grounds to search for cored dark matter profiles, for example 
via strong lens modelling \citep{zitrin:2012:inner, limousin:2022:lensing}, the internal kinematics of the brightest cluster galaxy (BCG; \citealt{kelson:2002}), weak lensing, and combinations of these probes \citep{newman:2013}. Evidence for both cored and cuspy cluster core profiles has been discovered using these methods \citep{biviano:2023}, with the source of the measured coring not yet clearly attributable to baryonic feedback or dark matter particle interactions alone. This uncertainty is due in part to the fact that baryonic effects have been demonstrated to both increase DM density slopes in the cores of clusters via adiabatic contraction \citep{Schaller:2015:corebaryons}, and decrease these slopes via dynamical friction \citep{inoue:2011,martizzi:2012:frictionandfeedback}, galaxy formation processes \citep{laporte:2012} and active galactic nucleus feedback \citep{ragone:2012:agnfeedback, peirani:2017:agnfeedback}.

The bulk motion of the BCG also serves as a promising probe of the cluster core potential. In a simplified picture one may use the BCG, typically found in the core of a galaxy cluster, as a test particle whose motion reveals the shape of the underlying potential set by the DM. In a relaxed cluster, the BCG is expected to rest close to the cluster core and oscillate on small length scales; the amplitude of oscillation relative to the potential minimum then reflects the inner slope of the density profile and therefore potentially the dark matter particle physics. This probe has been referred to as ``BCG wobbling'' (see \citealt{harvey:2017,harvey:2019}). Since in real observations only a snapshot of each cluster is available (due to dynamical timescales being much larger than those relevant to observations), a statistical approach is taken whereby the distribution of observed offsets between BCGs and their respective cluster potential minima is understood to encode the shape of the potential dictating the oscillations. 

Alternatively, active mergers of galaxy clusters represent a similar but distinct testing ground for SIDM physics, in particular by considering an effective drag force on the DM halos of two merging clusters, which would result in an offset between either the centroid and/or peak of the DM and assumed collisionless stellar component of each merger member. Whether an offset is expected to develop between the centroid and/or peak is dependent on the macroscopic model of the action of the drag force \citep{markevitch:2004, harvey:2014}, in addition to aspects of the microscopic interactions such as the scattering rate and typical scattering angles \citep{kahlheofer:2013}. These models are summarized and tested by \cite{kim:2017}, in which the development of BCG-DM offsets are studied in the context of staged equal-mass mergers in DM-only simulations, with stars treated as tracer particles. These mergers are examined both in CDM and SIDM, with varying self-interaction cross section per unit dark matter mass $\sigma_{\rm{SI}}/m \in [0, 1, 3, 10]\,\rm{cm^2\,g^{-1}}$ where zero is the CDM case. Here it is observed that the BCG-DM offset oscillates with very little damping post-merger, with typical offset sizes monotonically increasing with self-interaction cross section. At $1\,\rm{cm^2\,g^{-1}}$ offsets are measured to be on the order of $\sim 20\,\rm{kpc}$, whereas at $10\,\rm{cm^2\,g^{-1}}$ offsets are $\sim 100\,\rm{kpc}$.  
Note that since typical dark matter particle velocities increase with cluster mass, and velocity-dependent dark matter self interaction models are preferred \citep{kaplinghat:2016:veldepsidm, sagunski:2021}, these studies which consider velocity-independent dark matter self interaction on the cluster scale should be understood as probing the high-velocity interaction cross section. 

In each case, the expectation for CDM is that there are zero offsets between bound DM and stellar distributions, with the assumption that they are both effectively collisionless. A precise quantification of ``zero'' however has not yet been fully established, and will be necessary to make the statement that any observation or set thereof is consistent with SIDM offset distributions at some self-interaction cross section and simultaneously \textit{inconsistent} with CDM predictions. In \cite{kim:2017} the CDM staged mergers exhibit oscillation amplitudes on the order of $\sim 10\,\rm{kpc}$, which, depending on the systematics of observations, may be indistinguishable from the $\sim 20\,\rm{kpc}$ offsets of the $\sigma_{\rm{SI}}/m = 1\,\rm{cm^2\,g^{-1}}$ dark matter model. Current observational constraints place upper limits on the self-interaction cross section on the order of $\sim 1\,\rm{cm^2\,g^{-1}}$ \citep{randall:2008:bullet} or even $\sigma_{\rm{SI}}/m < 0.19 \,\rm{cm^2\,g^{-1}}$ at 95\% confidence \citep{andrade:2021:sidmcrosssection}. For such a cross section, the SIDM oscillation amplitude is likely to be indistinguishable from that of CDM when viewed in the context of these staged merger results. This highlights the necessity for increased precision in the understanding of CDM offsets.   

The relaxed ``wobbling'' scenario has been studied in \cite{harvey:2017} by comparing BCG offsets in observations of relaxed clusters to oscillation amplitudes in the BAHAMAS suite of CDM simulations \citep{mccarthy:2017:bahamas}. The observed offsets are calculated as the on-sky distance between the positions of the BCGs and strong lensing centers, used as proxies for the potential minimum or dark matter center \citep{medezinski:2013:lensing, markevitch:2004, zitrin:2013:lensing}. Here it was found that offsets measured via mimicking wobbles analytically and performing mock strong lensing analyses on the simulation data were too small to be consistent with the observations, suggesting some degree of coring and thus self-interaction in the observations. This was developed into a constraint on the self-interaction cross section of DM in \cite{harvey:2019} using offset distributions in cosmological SIDM simulations and making ansatzes about the convergence of median offsets below the gravitational softening length, finding that the observational data used in \cite{harvey:2017} is consistent with CDM at the $1.5\sigma$ level and prefers a cross section of $\sigma_{\rm{SI}}/m < 0.39\,\rm{cm^2\,g^{-1}}$ at the 95\% confidence level. 

A hallmark of offset measurements made in relaxed clusters in simulations is that they are overwhelmingly below the gravitational softening length, and thus constitute measurements in the unconverged regime. The gravitational softening introduces an artificial coring independent of and degenerate with the coring due to self-interactions of the DM, but is necessary to avoid issues such as shot noise in the mass distribution unphysically affecting small-scale dynamics \citep{ludlow:2019}. Extrapolating below the softening length or even operating on scales similar to the softening length is known to produce untrustworthy results \citep{power:2003:convergence}, suggesting that to make reliable claims about dynamics and thus DM particle physics on scales of $\sim1-10\,\rm{kpc}$, simulations with sufficiently small softening lengths\footnote{Note that the softening length cannot be arbitrarily decreased, and appropriate choices for this parameter depend on other factors such as the 2-body relaxation time set primarily by particle \textit{number}. See \cite{ludlow:2019} for a detailed discussion.} should be utilized, and the convergence of offsets with respect to this parameter should be well established.  

In this paper, we investigate the offsets between the BCGs of galaxy clusters and various measures of their underlying oscillation centers across several cold dark matter cosmological hydrodynamical simulations. We make no attempt to distinguish between relaxed and unrelaxed clusters, instead aiming to produce a distribution of offsets over a whole cosmological volume, enabling comparisons with observations or simulation data which do not depend on relaxation criteria. The remainder of the paper is structured as follows: In Sec.~\ref{sec:data} we outline the simulations used to calculate offset distributions. In Sec.~\ref{sec:massmaps} we describe the procedure by which we generate mass maps which can be used to extract centroids for each particle species in these simulations. In Sec.~\ref{sec:centroid_identification} we define various centroid identification methods to be applied to the mass maps, including methods relevant to real observations. We discuss the offset distribution of each simulation considered, and the dependence of this distribution on centroid identification, redshift, mass, baryonic physics and various nuisance parameter choices in Sec.~\ref{sec:results}, and summarize these results in Sec.~\ref{sec:conclusions}.

\begin{deluxetable*}{lcccccccc}
\tablecaption{\normalfont\raggedright Physical parameters of the simulations considered in this article. Parameters are (from left to right): the simulation identifier, length of one side of the cubic simulation box, $N_{\rm{part}}$ is both the initial number of gas particles (some of which will be converted to stellar particles as time evolves), and also the number of DM particles (which remains unchanged), initial baryonic mass for moving-mesh code $M_{\rm{b}}$, dark matter particle mass $M_{\rm{DM}}$, gravitational softening length of the dark matter and star particles $\epsilon_{\rm{DM,\star}}$ at $z=0$, minimum gravitational softening length of the adaptive gas softening $\epsilon_{\rm{gas,min}}$ in comoving coordinates, number of friends-of-friends identified groups with mass above $10^{14}\,\msun$ at $z=0$, and number of friends-of-friends identified groups with mass above $10^{15}\,\msun$ at $z=0$.\label{tab:simulations}}
\tablehead{
\colhead{Simulation} & 
\colhead{\begin{tabular}[c]{@{}c@{}}Box Size\\ {[}cMpc$\,h^{-1}${]}\end{tabular}} &  
\colhead{$N_{\rm{part}}$} & 
\colhead{\begin{tabular}[c]{@{}c@{}}$M_{\rm{b}}$\\ $[\msun\,$$h^{-1}]$\end{tabular}} & 
\colhead{\begin{tabular}[c]{@{}c@{}}$M_{\rm{DM}}$\\ $[\msun\,$$h^{-1}]$\end{tabular}} & 
\colhead{\begin{tabular}[c]{@{}c@{}}$\epsilon_{\rm{DM, \star}}$\\ {[}kpc$\,h^{-1}${]}\end{tabular}} &
\colhead{\begin{tabular}[c]{@{}c@{}}$\epsilon_{\rm{gas,min}}$\\ {[}ckpc$\,h^{-1}${]}\end{tabular}} &
\colhead{\begin{tabular}[c]{@{}c@{}}$N_{\rm{groups}}$\\ $>10^{14}\,\msun$\end{tabular}} & 
\colhead{\begin{tabular}[c]{@{}c@{}}$N_{\rm{groups}}$\\ $>10^{15}\,\msun$\end{tabular}}\vspace{0.3em}
}
\startdata
IllustrisTNG300-1 & 205 & $2500^3$ & $7.6\e{6}$ & $4.0\e{7}$ & 1 & 0.25 & 460 & 6 \\
IllustrisTNG300-2 & 205 & $1250^3$ & $5.9\e{7}$ & $3.2\e{8}$ & 2 & 0.5 & 458 & 6 \\
IllustrisTNG300-3 & 205 & $625^3$ & $4.8\e{8}$ & $2.5\e{9}$ & 4 & 1 & 441 & 6 \\
MillenniumTNG740 & 500 & $4320^3$ & $2.0\e{7}$ & $1.3\e{8}$ & 2.5 & 0.25 & 6664 & 98 \\
MillenniumTNG185 & 125 & $1080^3$ & $2.0\e{7}$ & $1.3\e{8}$ & 2.5 & 0.25 & 98 & 1 \\
BAHAMAS & 400 & $1024^3$ & $8.1\e{8}$ & $4.5\e{9}$ & 4 & 4 & 2624 & 30
\enddata
\end{deluxetable*}

\section{Data}\label{sec:data}
We make use of three modern suites of cosmological simulations; IllustrisTNG (TNG; \citealt{nelson2021:illustristng}), MilleniumTNG (MTNG; \citealt{pakmor:2023:mtng}), and BAHAMAS \citep{mccarthy:2017:bahamas}. By analyzing offsets in the context of varying baryonic physics models and simulation parameters such as mass resolution, box size, and gravitational softening length, stronger statements can be made about the offset distribution prediction of CDM in simulations and the convergence of offset distributions with respect to various simulation nuisance parameters. A comparison of the simulation parameters for the data used from each suite can be found in Table \ref{tab:simulations}.  In each simulation we utilize only a minimum mass cut of $M_{200,\rm{mean}} > 10^{14}\,\msun$, where $M_{200,\rm{mean}}$ is the mass enclosed within $R_{200,\rm{mean}}$, defined as the radius within which the mean mass density is 200 times the mean matter density of the universe. We use this mass scale to match clusters used to infer offset measurements in the past, such as the bullet cluster \citep{randall:2008:bullet}, baby bullet \citep{bradac:2008:babybullet}, Abell 3827 \citep{williams:2011:abell3827} and El Gordo cluster \citep{jee:2014:elgordo}, which occupy masses from $\sim 10^{14}-10^{15}\,\msun$. See \cite{ng:2017} for a detailed comparison of existing observations. 

Since the WMAP 9-yr cosmology cosmology used in the BAHAMAS simulations (\citealt{bennett2013:wmap9yr}, 
$\Omega_{\rm{m}} = \Omega_{\rm{CDM}} + \Omega_{\rm{b}} = 0.2793,\: \Omega_{\rm{b}} = 0.0463,\: \Omega_\Lambda = 0.7207,\: \sigma_8 = 0.812,\:n_s = 0.972,\: h = 0.700$) is inconsistent with that used in both TNG and MTNG (\citealt{planck2016}; $\Omega_{\rm{m}} = \Omega_{\rm{CDM}} + \Omega_{\rm{b}} = 0.3089,\: \Omega_{\rm{b}} = 0.0486,\: \Omega_\Lambda = 0.6911,\: \sigma_8 = 0.8159,\:n_s = 0.9667,\: h = 0.6774$), we choose to analyze the data from each simulation with the corresponding cosmology, and it should be understood that there is a minor degree of incompatibility in all following comparisons, but we expect this to have no bearing on the measurements considered here. 

\subsection{IllustrisTNG}
The IllustrisTNG project (TNG; \citealt{springel:2017:ITNG, pillepich:2017:ITNG, marinacci:2018:ITNG, naiman:2018:ITNG, nelson2021:illustristng}) is a suite of cosmological hydrodynamical simulations spanning various simulation box sizes, mass resolutions, and physics models based on the \textsc{AREPO} code \citep{springel:2010:arepo}. This project succeeds the Illustris project \citep{vogelsberger:2014:illustris, genel:2014:illustris, vogelsberger:2014:illustrisgalaxies, sijacki:2015:illustris}, improving upon the galaxy formation model \citep{vogelsberger:2013:illustrismodel} and expanding the scope to larger volumes and higher resolutions. The TNG simulation suite consists of TNG50, TNG100 and TNG300, characterized by their varying simulation box sizes of $35$, $75$ and $205\, \rm{cMpc}$$\,h^{-1}$ respectively. 

We focus on TNG300 in this paper as we are interested in examining the largest galaxy clusters,  and this series of simulations admits the largest mass scales in TNG. TNG300 contains 3 full-physics runs in a $205\,\rm{\,cMpc}$$\,h^{-1}$ box named TNG300-1, TNG300-2 and TNG300-3, and their DM-only counterparts TNG300-1-Dark, TNG300-2-Dark, and TNG300-3-Dark, which share the box size and softening lengths of their corresponding full physics runs (see Table \ref{tab:simulations}) but at slightly lower mass resolutions of $7\e{7}$, $3.8\e{8}$ and $3\e{9}\,\msun$$\,h^{-1}$ than the full-physics counterparts respectively. This collection of mass scales, softening scales and physics models permit detailed resolution studies and the investigation of the effect of baryonic physics on the galaxies and galaxy clustering properties, with these simulations being used extensively in studies of galaxy clusters and their environments: eg. cool core formation in galaxy clusters \citep{barnes:2018:coolcore, barnes:2019:coolcore}, supermassive black hole feedback \citep{weinberger:2018:smbh}, and gas stripping dense environments in galaxy clusters \citep{yun:2019:jellyfish}. 

Groups in IllustrisTNG are identified via a friends-of-friends algorithm with a linking length $b=0.2$ in units of the mean interparticle separation 
\citep{springel:2001} with substructure identified via \textsc{subfind} \citep{dolag:2009:subfind}. This results in a total of $\sim450$ groups above $M_{200,\rm{mean}} = 10^{14}\,\msun$ in TNG300 at $z=0$ and 6 above $10^{15}\,\msun$ which are considered in this article, as in Table \ref{tab:simulations}. The original Illustris simulations have been used to study DM-BCG offsets and their dependence on centroid identification methods \citep{ng:2017}, but a detailed study of BCG offsets has not previously been conducted in IllustrisTNG.

\subsection{MillenniumTNG}
The MillenniumTNG project (MTNG; \citealt{hernandez:2023:mtng, pakmor:2023:mtng, barrera:2023:mtng}) is the successor to the dark matter-only Millennium simulation \citep{springel:2005:millennium}, the first 10 billion particle simulation in a $500\,\rm{Mpc}$$\,h^{-1}$ box. MillenniumTNG combines the volume of Millennium with the hydrodynamical modelling\footnote{With some small fixes in addition to changes relative to the TNG model, implemented to save memory in the much larger box \citep{pakmor:2023:mtng}.} developed for the IllustrisTNG project \citep{weinberger:2016:TNGgalaxyformation, pillepich:2017:TNGgalaxyformation} and is run using an improved version of the AREPO code \citep{pakmor:2016:arepo, weinberger:2020:arepo}. The MillenniumTNG simulations involve high time resolution merger trees and direct light cone output, enabling detailed cosmological inference via comparisons to large scale surveys \citep{contreras:2023:mtnglss, kannan:2023:mtnglss, bose:2023:mtnglss} and studies of the impact of baryons and massive neutrinos on weak lensing observations \citep{ferlito:2023:mtnglensing}. The MillenniumTNG suite consists of numerous dark matter-only runs at various resolutions, two hydrodynamical runs named MTNG740 and MTNG185 in analogy with the IllustrisTNG naming scheme, and a collection of dark matter + massive neutrino runs over a range of resolutions and neutrino masses. In this article we consider only the hydrodynamical runs.

The significantly larger box size of MTNG enables the study of much more massive clusters, with $98$ groups above $M_{200,\rm{mean}}= 10^{15}\,\msun{}$ at $z=0$ in comparison to the $6$ of TNG. In the context of DM-BCG offsets, this permits an analysis of offset distributions at various mass cuts relevant to strong lensing observations while maintaining sufficient statistics to study the tail of the distribution, a feature necessary to quantify in order to perform hypothesis tests using real observations. The dependence of offset distributions on mass cuts is discussed in Section \ref{sec:results}. 
The two hydrodynamical simulation runs in MTNG possess significantly different box sizes but identical gravitational softening length and mass resolution (see Table \ref{tab:simulations}). The motivation for these parameter choices is a study of box size effects on the matter power spectrum and the halo mass function \citep{hernandez:2023:mtng}. While there is no expectation that the offset distribution should be meaningfully influenced by the box size (other than the aforementioned potential dependence of the distribution on mass), we still make use of this data to perform a consistency check in Sec.~\ref{sec:results}.

\subsection{BAHAMAS}
The BAHAMAS suite of cosmological hydrodynamical simulations \citep{mccarthy:2017:bahamas, robertson2019:bahamas} has been a valuable asset in testing BCG-DM offsets, both in the CDM \citep{harvey:2017} and SIDM paradigms \citep{harvey:2019}. BAHAMAS (BAryons and HAloes of MAssive Systems) aims to study observables of large scale structure cosmology, and is based upon a version of \textsc{gadget-3} (last described in \citealt{springel:2005:gadget}) which is extended with subgrid physics including radiative heating/cooling, chemical evolution and both stelllar and active galactic nucleus feedback. These feedback prescriptions were calibrated to reproduce the present-day galaxy stellar mass function and hot gas fractions of clusters and groups, and the simulations have been successful in reproducing observations of stellar mass fractions in central galaxies, satellite galaxy dynamics, the local stellar mass autocorrelation function, the X-ray and SZ scalings of local groups and clusters, and black hole mass to velocity dispersion and stellar mass relations, among others. 

The BAHAMAS simulations use a $400\,\rm{cMpc}$$\,h^{-1}$ box following cosmo-OWLS \citep{lebrun:2014:cosmoowls, mccarty:2014:cosmoowls} and masses for the DM and baryonic particles a factor of $\sim 100$ larger than the highest resolution simulation considered here, IllustrisTNG300-1. As in Table \ref{tab:simulations}, the gravitational softening length for the DM and stars in this simulation is identical (other than a small disagreement due to chosen cosmology) to that of IllustrisTNG300-3 at $\epsilon_{\rm{DM, \star}} = 4\,\rm{kpc}$$\,h^{-1}$. This enables a comparison of offset measurements at a given softening length on similar mass scales, but with a different cosmology, feedback model and simulation code. For example the hydro method of BAHAMAS is based upon the smoothed particle hydrodynamics of \textsc{gadget-3} whereas the hydro method utilized in IllustrisTNG is the moving voronoi mesh of \textsc{AREPO}. This provides a valuable opportunity to establish whether the findings of the offset measurements in this paper depend on some specific choice in the modelling which differs between TNG and BAHAMAS, or if despite the differences in simulation methodology the offset distributions are consistent.

\section{Mass Maps}\label{sec:massmaps}
This paper uses particle data from N-body simulations to produce 2D mass maps of galaxy clusters. These mass maps will be used to measure the offset between the centroid of the stellar matter and measures of the oscillation center such as the potential minimum or centroids of the DM or gas distributions, with the purpose of illuminating the shape of the potential at the center of galaxy clusters. In this section, we describe the process by which we transform the particle position data of a galaxy cluster in an N-body simulation to a mock pixel image of the mass column density, discretized into pixels.

A mass map is a 2D function on the plane of the sky which represents the column mass density (units mass per area or per pixel). In order to make a mass map from simulation data which can be meaningfully compared to similar measurements in real observations, it is necessary to calculate the projected mass column density over a grid of pixels. In each pixel the density is calculated by smoothing each particle close to the line of sight with a kernel, and adding the contributions from all such kernels at the position of the pixel. The length of the line of sight is chosen to be $10\,\rm{Mpc}$, centered on the most bound particle in the cluster: this ensures the entire cluster is included in the column density without erroneously including contributions from nearby, unbound matter. To ensure this choice does not bias our results, we test offset distributions in IllustrisTNG300-1 at various line-of-sight depths in Sec.~\ref{sec:results}.

To perform the smoothing, we use the \textsc{SWIFTsimIO} projection module with the subsampled backend \citep{borrow2020, borrow2021}, which smooths particles to a fixed resolution grid using the Wendland-C2 kernel \citep{wendland:1995, dehnen:2012} with $\gamma = 1.936$. We choose this method for smoothing as it ensures well-converged results for the mass maps even at low pixel resolutions. To calculate the smoothing lengths for the star and DM particles, we use the \textsc{SWIFTsimIO} \texttt{generate\_smoothing\_lengths} method and for the gas we calculate the smoothing length for a Voronoi cell $i$ as $h_i = (m_i/V_i)^{1/3}$ where $m_i$ and $V_i$ are the total mass and volume of the Voronoi cell, respectively. 

For each galaxy cluster of interest, we project the DM, stars and gas particles onto a grid with pixels of side length $0.25\,\rm{kpc}$, centered on the position of the most bound particle. We choose this pixel size to be smaller than the physical size of Hubble Space Telescope pixels for objects at redshifts $z\sim 0.5$, which is approximately $0.1'' \sim 0.6\,\rm{kpc}$. The zoom box is chosen to have a side length of $50\,\rm{kpc}$ for the higher-resolution simulations (IllustrisTNG300-1, IllustrisTNG300-2 and MillenniumTNG), and a side length of $100\,\rm{kpc}$ otherwise. The $50\,\rm{kpc}$ figure is chosen to be as small as possible while still containing sufficient shape information for reliable centering, since larger fields of view can produce larger offsets. See Sec.~\ref{sec:hyperparameters} for tests of the impact field of view and pixel size have on the offset distribution. For the lower resolution simulations, the $50\,\rm{kpc}$ mass maps are too noisy for reliable center identification, so we double the box width to gain constraining power. For the large scale measurements described in Sec.~\ref{sec:centroid_identification} (lensing and large scale gas) we use a $300\,\rm{kpc}$ zoom box at the same physical pixel size, chosen to have half-width approximately twice the Einstein radius of clusters at this mass scale (see Sec.~\ref{sec:isophote}). Example mass maps for the highest-resolution simulation considered here, IllustrisTNG300-1, are shown in Fig.~\ref{fig:demogrid} along with the ellipses and various centroid identification methods described in the next section.

In Fig.~\ref{fig:demogrid} and the remainder of this paper, we use the terminology ``convergence radius'', a parameter which varies in definition  \citep{power:2003:convergence, bosch:2018:convergence} but also potentially from halo to halo in a simulation once a definition is fixed. Following \cite{ludlow:2019}, the convergence radius may be estimated as $r_{\rm c} = 0.055\,l$ where $l$ is the mean comoving interparticle separation of dark matter, $L_{\rm{box}}/N_{\rm{part}}^{1/3}$, with $L_{\rm{box}}$ the simulation box size and both parameters available in Table \ref{tab:simulations} for the simulations considered here. This considers the two-body relaxation between dark matter particles as the major mechanism affecting the convergence of halo central density profiles. Meanwhile, gravitational softening can also naturally ``smooth'' the matter distribution at a scale smaller than a few times $\epsilon_{\rm{DM,\star}}$. For simulations studied here, $\epsilon_{\rm{DM,\star}}$ values were specifically chosen to be of the same order as (and slightly smaller than) $r_{\rm c}$. Therefore, for simplicity in the rest part of this paper, we choose to take a conservative estimate of $r_c = 2.8\epsilon_{\rm{DM,\star}}$, motivated by the ``spline softening length" used by the \texttt{GADGET} code as the length scale above which pairwise forces become exactly Newtonian~\citep{ludlow:2019}, but note that this choice is not impactful for the present analysis. 

For the DM distribution, the mass per area is a reasonable metric for extracting centroids comparable to observations, however in the case of baryonic matter it is not stellar or gas mass that contributes to observations directly, but rather the luminosity. Basing these conclusions on mass maps is equivalent to a ``mass traces light'' assumption, not to be confused with the ``light traces mass'' assumption used to infer the position of DM or trace the potential using gas. The quality of this assumption can be tested by creating mock images of the baryonic matter and assessing the difference in centroid between this mock image and that obtained from a mass map. However for the galaxy population considered here, that is giant ellipticals in the cores of clusters, the observed ages and thus mass to light ratios will be highly uniform, with little scatter between galaxies and within galaxies themselves \citep{groenwald:2014}. As a result, we identify centroids directly from the mass maps and leave a detailed treatment of mock images to future work. 

\section{Centroid Identification}\label{sec:centroid_identification}
To measure offsets between the stellar mass distribution and a measure of the oscillation center, one must first establish a centroid for the stars and potentially other species like the DM or gas. In this section, we outline how we identify the position of the BCG in addition to various measures of the oscillation center. These are centroids obtained via the position of the most-bound particle in a cluster, source extraction via thresholding, and the centers of isophotes of cluster DM mass maps on scales relevant to cluster strong lensing. One could also consider the usage of the cluster center of mass as a proxy for an oscillation center, however this probe is too sensitive to large-scale structure which is unrelated to the small-scale oscillation of the BCG in the core.  

\subsection{Most Bound Particle}
\label{sec:potentialmin}
One method of approximating the oscillation center is using the potential minimum, identified by the position of the particle in the group of interest with the lowest gravitational potential energy\footnote{This position is also used as the ``group position'' field in the \textsc{AREPO}-based simulations \citep{springel:2010:arepo, weinberger:2020:arepo} and the center of the mass maps in Sec.~\ref{sec:massmaps}.}. This centroid identification method has the advantage of being easily interpretable via the simplified picture of a BCG oscillating in a fixed potential, however since gravity is improperly resolved below the softening length in a simulation, this position is uncertain on a scale close to the softening length. The simplified picture of a fixed potential in which the BCG test mass oscillates ignores the contribution to the potential of the BCG itself, whose size ($\sim 100\,\rm{kpc}$) is significantly larger than the length scale of oscillations ($\sim 10\,\rm{kpc}$). It is possible that this combined potential minimum oscillates \textit{with} the BCG but at lower amplitude, perhaps even with discontinuities due to the most bound particle changing at each time step, artificially reducing offset measurements. Whether these offsets are in fact systematically smaller than other centroid identification methods will be examined in Sec.~\ref{sec:results}. All references to an ``oscillation center'' in this paper should therefore be treated with some nuance, as the underlying potential is time-dependent.

\begin{figure*}
    \centering
    \includegraphics[width=\linewidth]{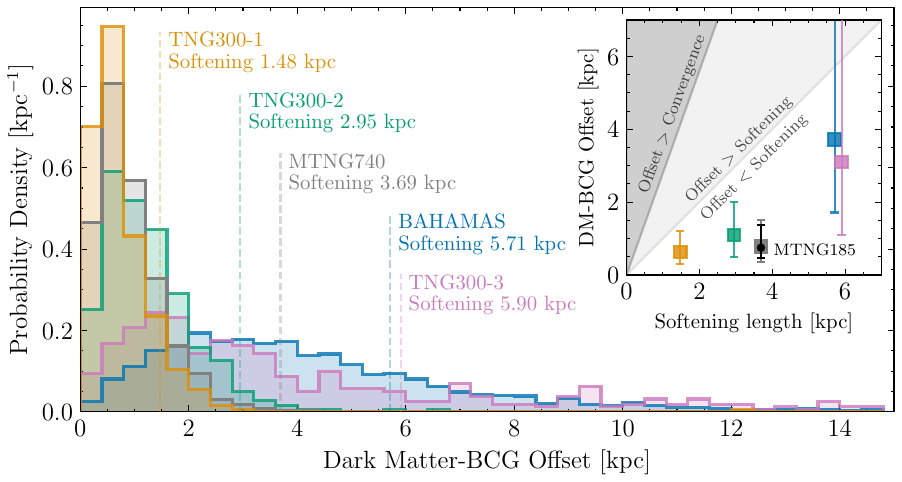}
    \caption{Dark matter - brightest cluster galaxy centroid offsets in the cores of galaxy clusters above $10^{14}\,\msun$ in the IllustrisTNG, MillenniumTNG and BAHAMAS cold dark matter cosmological hydrodynamical simulations. Offsets are measured by finding the dark matter and stellar density centroids using SourceExtractor on projected mass density maps smoothed from the raw particle data using a smoothed particle hydrodynamics interpolation scheme. (\emph{Inset}) the median and $68\%$ bounds of the offset distribution for each simulation, including an additional data point for the MillenniumTNG185 run, which possesses the same gravitational softening length but a smaller box than its MillenniumTNG740 counterpart. Convergence refers to the convergence length, calculated as $2.8\epsilon_{\rm{DM}}$ where $\epsilon_{\rm{DM}}$ is the gravitational softening length (see text).
    \label{fig:offset_comparison}}
\end{figure*}

\subsection{Centroid Fitting}\label{sec:centroid_fitting}
The offsets between centroids of different particle species, such as DM-stars or gas-stars offsets, can also be used as a proxy for the shape of the potential. To locate faithful centroids of these distributions, one can utilize the mass maps of Sec.~\ref{sec:massmaps}. A centroid can be extracted from these maps by various means, for example by assuming a profile shape and fitting the profile using a minimization or posterior sampling scheme. One can also extract centroids without assuming a shape for the profile using an algorithm like shrinking spheres in 3D (eg. \citealt{schaller:2015}), shrinking apertures in 2D, (eg. \citealt{ng:2017}), or by thresholding, whereby an underlying statistical variation is identified as a ``background'', and any pixel a certain threshold above this background is classified as belonging to an object (eg \citealt{harvey:2019}). In the thresholding approach, the object pixels can then be divided into different sources, if present. A standard software package which performs this thresholding procedure is SourceExtractor (SE; \citealt{bertin:1998:sextractor, barbary:2016:sep}) and we choose to fit centroids of the DM, star and gas distributions using this method as opposed to profile fitting or convergent shrinking algorithms. This has the potential advantage of mimicking some systematics of the procedure applied to astronomical images to identify centroids of stellar and gas distributions, and we treat the DM in the same manner for consistency, noting that such a measurement would not be possible for real observations. This choice is consistent with the treatment of the stellar maps of \cite{harvey:2019}, in which the stellar centroid was obtained via thresholding with SourceExtractor and the dark matter centroid/oscillation center was taken as the potential minimum (Sec.~\ref{sec:potentialmin}). Centers of stellar mass maps in this article are \textit{always} identified using this thresholding procedure. 

After extracting sources in a mass map, we choose the center of the source with the highest ``flux'' as the centroid in our image, since the highest flux is very likely to be associated with the flux of the large-scale source including the center. The term ``flux'' is borrowed from the extraction methods of SourceExtractor, and should be understood here in the context of the mass traces light assumption. Examples of ellipses describing the center and orientation of such sources are shown in Fig.~\ref{fig:demogrid} in the ``DM'', ``stars'' and ``gas'' panels. 

We refer to the center of the highest-flux source in the stellar mass map as the position of the ``BCG", however it has been demonstrated \citep{Bosch:2005,Skibba:2011,Lange:2018} that $\sim 30\%$ of clusters have central galaxies which are not their brightest member galaxy. Therefore the offsets in this paper are most accurately described as offsets between some measure of the oscillation center and the brightest source within $\lesssim 25\,\rm{kpc}$ of the potential minimum (set by half the side length of the zoom box, and therefore this number is variable). This source may be the BCG or a central galaxy which is not the brightest member, but for simplicity we maintain the terminology of a ``BCG offset". Note that various other stellar centroid measures exist, for example in \cite{ng:2017}, stellar centroids such as the peak of the \textit{i}-band flux, an \textit{i}-band luminosity-weighted centroid of many member galaxies, and the peaks of heavily smoothed kernel density estimates (KDEs) of the galaxy luminosities were each considered, with the peak of the $i$-band flux exhibiting the least scatter. 

The SE object identification algorithm relies on a number of nuisance parameter choices, for example the kernel size used to determine the ``background'' noise level, and the threshold above this level for a detection to be made. This threshold and other nuisance parameter choices are not expected to have a significant impact on the determination of the position of the central source, however we test the stability of the offset measurement to these choices in Sec.~\ref{sec:results}. When extracting sources and thus identifying centers, we choose background extraction parameters as a function of the pixel resolution of the mass map, denoted $N_{\rm{pix}} \times N_{\rm{pix}}$ as \texttt{bw=bh=$\lceil N_{\rm{pix}}/4 \rceil$,fw=$\lceil$bw$/10\rceil$,fh=$\lceil$bh$/10\rceil$} (where $\lceil x \rceil$ represents the ceiling function) and \texttt{threshold=$1.5$} which denote the widths and heights of the background boxes and filters of the SE detection algorithm, respectively, and the signal to noise threshold above which an object is considered detected.

\subsection{Area-Matched Isophote}\label{sec:isophote}
In observations, the oscillation center has been estimated using the center of the primary halo in a strong lensing model \citep{markevitch:2004, medezinski:2013:lensing, zitrin:2013:lensing}. Strong lens modelling is sensitive to the large-scale potential of the system set by the DM, and not necessarily to the precise shape of the core. The large-scale information can however be used to ``triangulate'' the center under the assumption that the DM mass distribution possesses at least two axes of symmetry on the sky. An example of such a distribution would be a normal distribution\footnote{The large-scale halos in strong lens modelling are typically modelled by other profiles such as generalized forms of the Navarro-Frenk-White profile or the pseudo-isothermal sphere \citep{navarro:1996, navarro:1997}.} in two dimensions $(x,y)$ with variance along each axis $\sigma_x$ and $\sigma_y$, not necessarily equal. This distribution has symmetry axes along $x$ and $y$ and its contours are ellipses with a shared and well-defined center, such that precise knowledge of an outer contour yields precise knowledge of the distribution center. More generally, the center can be identified even if the shape of the distribution is unknown, provided it possesses this symmetry. 

The shape of DM halos in cosmological simulations are typically described by triaxial ellipsoids \citep{allgood:2006}, and the projection of such a distribution onto the plane of the sky is again an ellipse, so we expect the deviation from the above assumption to be small for relaxed systems, and thus the lensing center to be a good approximation for the DM centroid. However, for systems undergoing oscillations, the small-scale oscillations may not propagate to larger scales on sufficiently short timescales, and can lead to an offset between the center of the small-scale information and the large-scale. In this case, the large-scale lensing center will be an inappropriate proxy for the oscillation center and thus lead to spuriously large offset measurements. This will be exacerbated in unrelaxed systems such as that shown in Fig.~\ref{fig:demogrid}, for which the strong lensing scale encompasses more than one large DM halo and the small scale centroid is significantly offset ($\gtrsim 20\,\rm{kpc}$) from the large scale DM halo center.

To mimic this feature, that is an estimation of the DM centroid on a scale to which lensing is sensitive, we introduce a new centroid identification method. We utilize the mass maps of Sec.~\ref{sec:massmaps} and extract the mass isophotes\footnote{A term typically used for photometry, but we treat the mass maps as analogous to images.} via \textsc{photutils} \citep{bradley:2023:photutils}. 
Unlike contours, the isophotes possess a well-defined center and yield an approximation of the center assuming elliptical symmetry, mimicking the assumptions of the lensing analysis without the requirement that the isophote ellipses share a center. To perform the isophote fitting, we use the \texttt{fit\_image} method of the \texttt{isophote.Ellipse} class, with parameters as follows: the minimum and maximum isophote semi-major axes are set by the mass map pixel resolution $N_{\rm{pix}}$ as $a_{\rm{min}} = 0.1 N_{\rm{pix}}$, $a_{\rm{max}}=0.7 N_{\rm{pix}}$. Note that $N_{\rm{pix}}=(50\,\rm{kpc})/(0.25\,\rm{kpc\,pixel^{-1}})=200$ for the typical $50\,\rm{kpc}$ mass map size considered here. We also use a minimum of 20 and maximum of 50 iterations of the fitting algorithm, and use lenient convergence parameters of \texttt{maxgerr=$1.0$, fflag=$0.8$} and \texttt{conver=$0.01$} to obtain estimates even when isophote identification is challenging due to the presence of substructure, for example. This lenience is justified by the lack of uncertainty in the mass map of the simulation, and the fact that dealing with regions of low signal to noise in the faint outskirts of images, for which parameters such as \texttt{maxgerr} are intended, is less relevant in this context.

Upon identifying the isophotes, we choose the isophote with the same on-sky area as a circle formed by the Einstein radius \citep{narayan:1997:lectures}. Instead of calculating an Einstein radius by assuming the cluster is a point mass, or by assuming a shape for the mass distribution such as the NFW profile or the isothermal sphere \citep{navarro:1996, navarro:1997} for which analytical Einstein radii have been calculated \citep{narayan:1997:lectures, dumet:2013:nfweinsteinradius}, we instead choose to estimate an approximate Einstein radius from observations. For sources at redshift $z_S\sim 2$ and 37 galaxy cluster lenses at $z_L\sim 0.5$ in a similar mass range to those clusters considered in this paper, \cite{sharon:2020} find a median Einstein angle of $\theta_E=10.8''$ which corresponds to a lens-plane radius of approximately $70\,\rm{kpc}$. We use this distance as the Einstein radius when selecting the area-matched isophote.

This Einstein area-matched isophote center will be subject to the same potential issues of miscentering between small and large scales as observations using strong lensing, but the precise nature of this effect will not be properly modelled here. In future work, we will examine mock lensing images of clusters in simulations and compare the offset distribution to that measured on smaller scales, in particular assessing if this effect leads to a discrepancy between observed offset distributions and those of simulations. This can be accomplished by assuming statistics of a background distribution of sources, or by utilizing light cones such as those output by MillenniumTNG which ensures self-consistency of the background distribution with the lens. 

\section{Results}\label{sec:results}
We now apply the centroid identification techniques of Sec.~\ref{sec:centroid_identification} to the mass maps of Sec.~\ref{sec:massmaps} to identify the distribution of offsets in each simulation suite, and investigate its dependence on gravitational softening length, centroid identification method, redshift and the mass range of clusters considered. We also briefly review the impact of the baryonic model in IllustrisTNG on the density profiles of the cluster halos, in particular comparing the coring of the profiles in the full-physics runs to those of the DM-only runs.

\subsection{Dependence on Softening Length}\label{sec:results_softening}
The gravitational softening length introduces a flattening of the gravitational potential below a scale $\epsilon_i$, which is identical for the dark matter and stellar particles, though note that the gas uses an adaptive softening length scheme with a minimum value $\epsilon_{\rm{gas,min}}$ in IllustrisTNG and MillenniumTNG, and a fixed softening length in BAHAMAS. This artificial coring could induce oscillations of the BCG degenerate with the signal of interest, that is the oscillation due to cored central potentials arising from DM self-interaction. We calculate the offset distribution in the three resolution levels of IllustrisTNG300, two box sizes with identical resolution of MillenniumTNG, and the single BAHAMAS resolution considered, where offsets are measured as the 2D on-sky distance from the SourceExtractor-identified DM centroid to the SourceExtractor-identified stellar centroid. The effect of the centroid identification method is investigated in Sec.~\ref{sec:results_centroid_choice}. 

In Fig.~\ref{fig:offset_comparison} we present these offset distributions across all simulations considered, as measured by SE centroids. 
Here it can be seen that the majority of the offset distributions are below the softening length scale in each case, and that the IllustrisTNG 300-3 and BAHAMAS distributions are consistent, while possessing almost identical softening lengths. There also exists a clear trend of decreasing offsets with decreasing softening length, which is the expectation in a physics model whose ``true'' offset distribution is consistent with zero, or at least smaller than the smallest softening length considered here, namely $1.48\,\rm{kpc}$ in IllustrisTNG 300-1. We also represent the median and 68th percentile offsets as a function of softening length in the inset panel of Fig.~\ref{fig:offset_comparison}. Here it is made clear that with centroids measured by SourceExtractor, the offset measurements are limited by the gravitational softening and convergence properties, consistent with the findings of \cite{schaller:2015}, \cite{ng:2017} and \cite{harvey:2019} in which offsets measured between the DM centroid (or potential minimum) and BCG are typically on the order of or below the softening length. 

The lower resolution simulations do exhibit some offsets above softening, but these may be due to resolution and zoom box nuisance parameters, as the relatively poor mass resolution necessitates a larger zoom box used to create the mass maps. As discussed in Sec.~\ref{sec:hyperparameters} the size of the zoom box positively correlates with the median offset, as measuring the center on a larger scale has the potential to deviate due to substructure. A small amount of offsets above softening are observed in all distributions, and can often be identified by visual inspection as caused by merging cores and bimodal mass distributions, or a failure of the source extraction to faithfully represent the large scale halo in rare cases. We leave a detailed study of the physical nature of large offsets to future work, focusing instead here on the limitations of CDM simulations and the rarity of offsets above softening. 

We also note that the two MillenniumTNG runs yield highly consistent offset distributions, despite the large difference in simulation box size and thus mass scale. This feature will be further tested with the mass cuts of Sec.~\ref{sec:mass_redshift}.

\subsection{Dependence on Centroid Identification}
\label{sec:results_centroid_choice}

\begin{figure}
    \centering
    \includegraphics[width=0.9\linewidth]{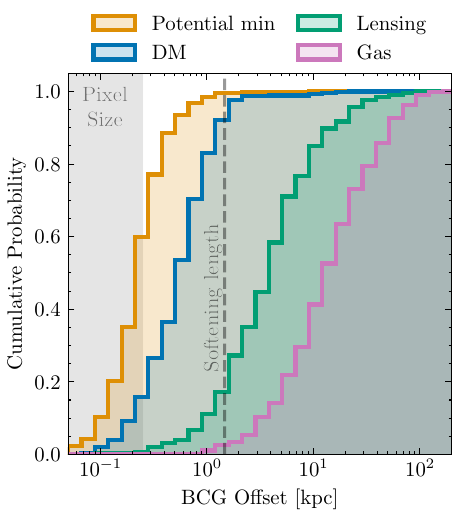}
    \caption{Comparison of offset distributions in IllustrisTNG300-1 as measured by the four oscillation center identification measures/proxies studied in this article. Each distribution is measured via the offsets between that method's centroid and the SourceExtractor-identified BCG position, applied to all groups with mass above $M_{200,\rm{mean}}=10^{14}\,\msun{}$ at $z=0$. Note that ``Gas'' refers to the large-scale gas measurement, and colors are roughly matched to Figure \ref{fig:demogrid}. The cumulative probability is the probability that an offset is measured as less than or equal to the corresponding $x$-axis value.}
    \label{fig:center_choice}
\end{figure}

The offset measurement relies fundamentally on the identification of two points, one center for the stellar matter (BCG), and another point which infers an oscillation center. We use SourceExtractor to extract the centroid of the stellar matter, mimicking the procedure in real observations, but there are multiple choices for estimating the oscillation center, such as the potential minimum, a SourceExtractor centroid for the DM, or the potential isophote method described in Sec.~\ref{sec:centroid_identification}. It has also been assumed (eg. \citealt{cross:2023}) that ``light traces mass'' (LTM), which is used as motivation to estimate the oscillation center using that of the gas in an indirect measurement. Together, this constitutes a set of four BCG offset measurements which we discuss in this Section.

In Fig.~\ref{fig:center_choice} we represent the offset distribution of IllustrisTNG300-1 as measured between SourceExtractor stellar centroids (``BCG'') and these four choices for the oscillation center. Here one can see that the smallest offsets are measured between the BCG positions and the potential minimum. This is followed by the BCG-DM offsets, with the majority of the offset distributions for these two methods measured on sub-softening scales. That the SE-SE offsets are approximately double the potential minimum-SE offsets on average can be seen as a twofold measurement of the same noise in source extraction, once in identifying the BCG and again in identifying the DM centroid. This could also be understood conversely by considering the potential minimum-SE offsets as a reduction of the more faithful SE-SE offset measurements. This can be understood in terms of the discussion of Sec.~\ref{sec:potentialmin}. The stability of the source extraction to the choice of background identification threshold is established in Fig.~\ref{fig:hyperparameter_tests}, demonstrating that in the interpretation of potential min-SE offsets being more faithful, the source extraction has some intrinsic noise for a given mass map that is unrelated to nuisance parameter choice. This noise however is on sub-softening scales, and so does not meaningfully bias any inference made using these simulation offset distributions.

\begin{figure*}
    \centering
    \includegraphics[width=0.9\linewidth]{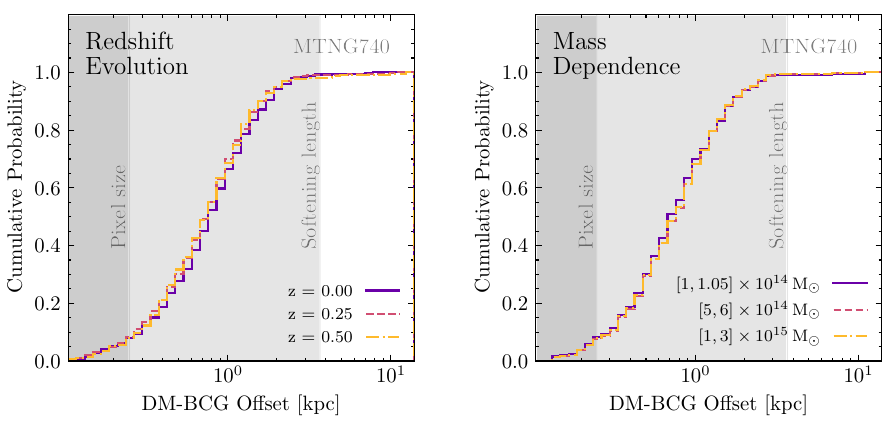}
    \caption{Variation in dark matter-brightest cluster galaxy offsets in MillenniumTNG740, as measured by centroid fitting with SourceExtractor, with the redshift (left) and mass range (right) of clusters considered. Offset distributions at varying redshift are obtained using offset measurements in all clusters above $M_{200,\rm{mean}} = 3\e{14}\,\msun{}$, the scale of mass distribution decreasing with increasing redshift. The cumulative probability is the probability that an offset is measured as less than or equal to the corresponding $x$-axis value. The number of clusters above this limit at each redshift is 1197 ($z=0$), 688 ($z=0.25$), and 311 ($z=0.5$). The mass dependence is assessed at $z=0$, and the number of groups in each mass bin are 438 ($[1,1.05]\e{14}\,\msun{}$), 600 ($[5,6]\e{14}\,\msun{}$), and 681 ($[1,3]\e{15}\,\msun{}$).}
    \label{fig:redshift_mass_dependence}
\end{figure*}

When the position of the DM centroid is measured via the area-matched isophote, the offset distribution extends to offsets roughly a factor of 10 larger than the potential minimum and SE-SE distributions, with roughly $80\%$ of the lensing scale offset distribution being measured above the softening length. The miscentering between scales leading to this discrepancy can be seen in the example of Fig.~\ref{fig:demogrid}, wherein the Einstein isophote center falls far outside of the convergence radius.

This hints at significant implications for lensing analyses, in particular those of a small number of systems. In such a case, it is difficult to make a statistically significant claim that the observations are inconsistent with the offset distribution of CDM, given that scale miscentering can increase offset measurements by an order of magnitude relative to more faithful measures. The total uncertainties on strong lensing centers relevant to these analyses are typically on the order of $\pm \sim 1''$ \citep{zitrin:2015:clash, harvey:2017}, corresponding to a $\pm5\,\rm{kpc}$ lens-plane distance. If the lensing center in this approximate picture can be offset from the underlying faithful DM center by up to $\sim 10\,\rm{kpc}$, as in Fig.~\ref{fig:center_choice}, then the relatively small uncertainties on the observed lensing center may lead to offset measurements falsely interpreted as large oscillation amplitudes, when in fact it is assumptions (eg. symmetry) which are breaking down. This highlights again the caution which much be taken when comparing offset distributions in simulations to those of observations, as not only are many of these offsets in an unconverged regime, but also systematics of the centroid measurement method can significantly and artificially inflate offsets to appear inconsistent with CDM. A full treatment of the offset distribution obtained via mock lensing analysis is necessary however to compare this feature directly to observations.

Finally, the large scale gas-BCG offsets are the largest of any centroid identification method, indicating that the gas is a poor tracer of the stellar, and therefore dark matter in the core. This can also be seen in Fig.~\ref{fig:demogrid}, as in the large-scale gas the centroid is offset from the potential minimum by $\sim 50\,\rm{kpc}$, while the small scale structure of the gas is highly clumpy, making a small scale measurement of the gas centroid only loosely correlated with the underlying potential. These findings are qualitatively in line with those of \cite{Seppi:2023}, in which the BCG-gas offsets were measured by various means in simulations (including TNG300-1) and in observations, as a means of estimating cluster disturbance. The BCG was defined in \cite{Seppi:2023} to lie at the potential minimum, which we demonstrate to be coincident above the softening scale with the BCG in Fig.~\ref{fig:center_choice}, and the gas center was computed via an emission measure weighted center of mass. The gas-BCG offsets in this paper are however systematically smaller by a factor of $\sim 2$, likely owing to the significant difference in centroid identification methodology. The scale of these offsets is also consistent with observations such as those of \cite{Lauer:2014} in which 433 gas-BCG offsets were observed, 15\% of which were above $100\,\rm{kpc}$. The offsets observed here are again slightly smaller than those observed at 15\% above $\sim 60\,\rm{kpc}$, but it should be noted that many observational effects are not modelled here. 

That the gas is an unreliable tracer of the potential minimum in simulations is a robust statement as the offsets measured are above the convergence radius of the underlying dark matter. Though it is possible that the gas physics is improperly resolved on small scales, the large scale gas centroid is more relevant to observations and is a poorer tracer of the potential minimum than even the small scale centroids. Thus inferences made using the gas as a tracer for the potential center, on both large and small scales, can lead to misinterpretations of the underlying dark matter physics.

\subsection{Dependence on Redshift and Mass Scale}\label{sec:mass_redshift}
The offset distribution can in principle depend on the redshift of the simulation snapshot examined, in that the distribution of cluster masses changes over cosmological time, as does the major merger rate \citep{fakhouri:2010, rodriguezgomez:2015:mergers}. We test the redshift evolution of the offset distribution using MillenniumTNG740 at redshifts of $z = 0, 0.25$ and $0.5$. We choose MillenniumTNG740 for this test as the redshift evolution is coupled with the mass cut tests, and those tests are more readily conducted with the larger mass scale and greater statistics of MTNG. We present these results in Fig.~\ref{fig:redshift_mass_dependence}, where it can be seen that the offset distribution at each redshift is entirely consistent, and furthermore the offset distribution in each case is almost entirely sub-softening. If these methods are primarily probing noise in the centroid identification method coupled to a lower bound set by the softening scale, as demonstrated by Fig.~\ref{fig:offset_comparison}, then it is not surprising that the offset distribution is not changed in physical coordinates with redshift, since the softening length is fixed at these redshifts in MillenniumTNG. 

Also in Fig.~\ref{fig:redshift_mass_dependence} are the $z=0$ offset distributions at various mass cuts in MTNG, with the goal of testing whether the offset distribution depends on the mass scale of clusters considered. Previous work in BAHAMAS suggests that the median offset should increase with cluster mass between $10^{14}$ and $10^{15}\,\msun{}$ \citep{harvey:2019} in SIDM runs, but that no mass trends are found in CDM. We test the CDM prediction over a slightly wider mass range, permitted by the larger box size of MTNG. We find no significant difference between offset distributions at $M\sim10^{14}\,\msun{}$ and $M \sim 10^{15}\,\msun{}$, in particular finding the vast majority of offsets below the softening length, consistent with the results of Fig.~\ref{fig:center_choice} in IllustrisTNG300-1. This finding is supported by the comparison between the MTNG740 and MTNG185 offset distributions in Fig.~\ref{fig:offset_comparison}, whereby despite the significantly different mass statistics, the offset distributions are highly consistent.

Understanding the oscillation dynamics in the merging and relaxed phases \citep{kim:2017} of a galaxy cluster on mass scales relevant to observations necessitates the study of high time-resolution simulation data of a merger involving a halo on the $10^{14}-10^{15}\,{\rm{M}}_\odot$ mass scale, with a softening length comparable to the high-resolution simulation suites considered here ($\lesssim 3\,\rm{kpc}$) and a full treatment of the baryonic physics. The snapshot output rate of all simulations considered here is insufficient to successfully resolve merger dynamics in time, as for a length scale of $100\,\rm{kpc}$ the crossing time of cluster members travelling at $\sim 10^3\,\rm{km\,s^{-1}}$ in the center of a cluster will be on the order of $\sim 100\,\rm{Myr}$. IllustrisTNG mini-snapshots are stored at $\sim100-200\,\rm{Myr}$ intervals, meaning that a factor of $10-100$ increase in snapshot output frequency would enable this analysis. Thus future works may study the evolution of offsets at this higher time resolution to understand both the merger and relaxed phases of oscillation, along with the faithfulness of centering methods in each phase. 

\begin{figure*}
    \centering
    \includegraphics[width=0.9\linewidth]{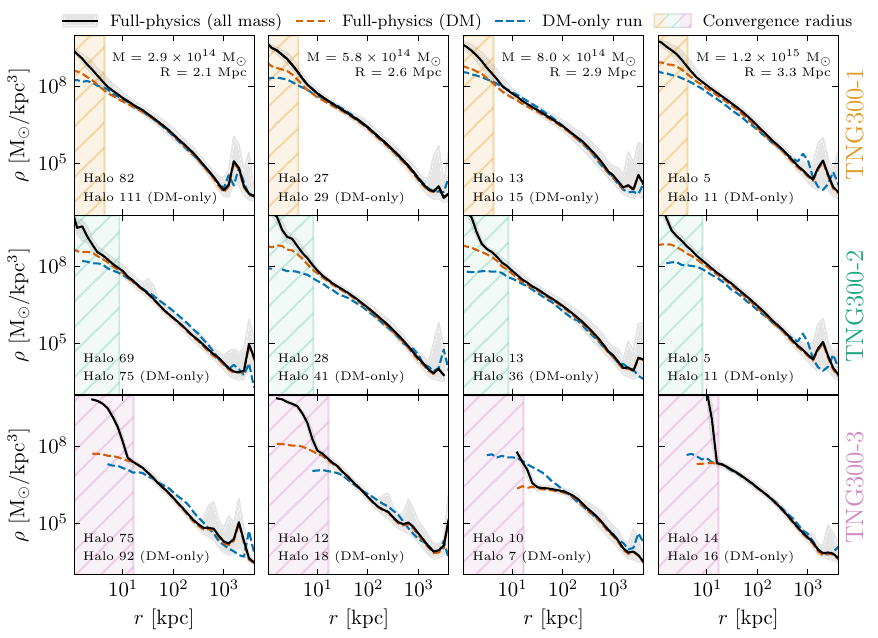}
    \caption{Density profiles of example halos in full-physics IllustrisTNG300 runs at three resolution levels, with the density profiles of their matched dark matter (DM)-only run counterparts. Shown are the total density profiles and DM density profiles in the full-physics run, and the DM density profiles in the DM-only run. Halos are also matched across simulations (along columns) in both mass and radius to below displayed precision, to demonstrate how similar halos look at each resolution level. Halo IDs in the group catalogs of the full-physics and DM-only runs are also provided. The convergence radius is calculated as $2.8\epsilon_{\rm{DM}}$ where $\epsilon_{\rm{DM}}$ is the gravitational softening length (see text). 
    }
    \label{fig:baryons_density_profiles}
\end{figure*}

\subsection{Dependence on Baryons}
We also test the influence of the baryonic physics model on the shape of cluster core density profiles. If the baryonic physics model contributes to a significant coring or cusping of the central density profile, this could lead to offset measurements which are more effective probes of the baryonic physics than the underlying DM self-interaction properties. In Fig.~\ref{fig:baryons_density_profiles} we examine the density profiles of IllustrisTNG300 at three resolution levels, and compare these density profiles to their counterpart halos in the DM-only run, where the full-physics to DM-only matching is described in \cite{lovell:2018, nelson2021:illustristng}. We also match halos according to their mass and radius across simulation resolutions to assess if any trend in cluster size exists, with the matching in both parameters at the $\lesssim$1\% level.

In Fig.~\ref{fig:baryons_density_profiles} it can be seen that the effect of the baryonic model on the DM density profiles (dotted lines) is to increase the density in the core, consistent with the adiabatic contraction caused by gas condensation and mass accretion seen in other simulations \citep[eg.][]{Schaller:2015:corebaryons}, but only below the convergence radius. Since these changes take place primarily below the convergence radius, robust claims about the impact of this baryonic model on offset measurements cannot be made. For example offsets are overwhelmingly measured below the softening scale, which may be in part due to increased central densities as a result of this contraction, but this cannot be concluded reliably from this data. It is also the case that no significant trend appears in cluster size, with the evidence of contraction being comparable at all mass levels shown.

One factor not explicitly considered here is the efficiency with which AGN feedback ejects particles on the resulting offset distributions. It may be beneficial to perform a systematic study of offsets in an isolated system with variable AGN feedback parameters, or to establish the offset distribution in a simulation with stronger feedback model such as SIMBA \citep{dave:2019:simba}, whose cluster mass distribution is similar in scale to that of MTNG185 due to its $147.1\,\rm{cMpc}$ box. Such a study is however outside the scope of this paper.

\begin{figure*}
    \centering
    \includegraphics[width=\linewidth]{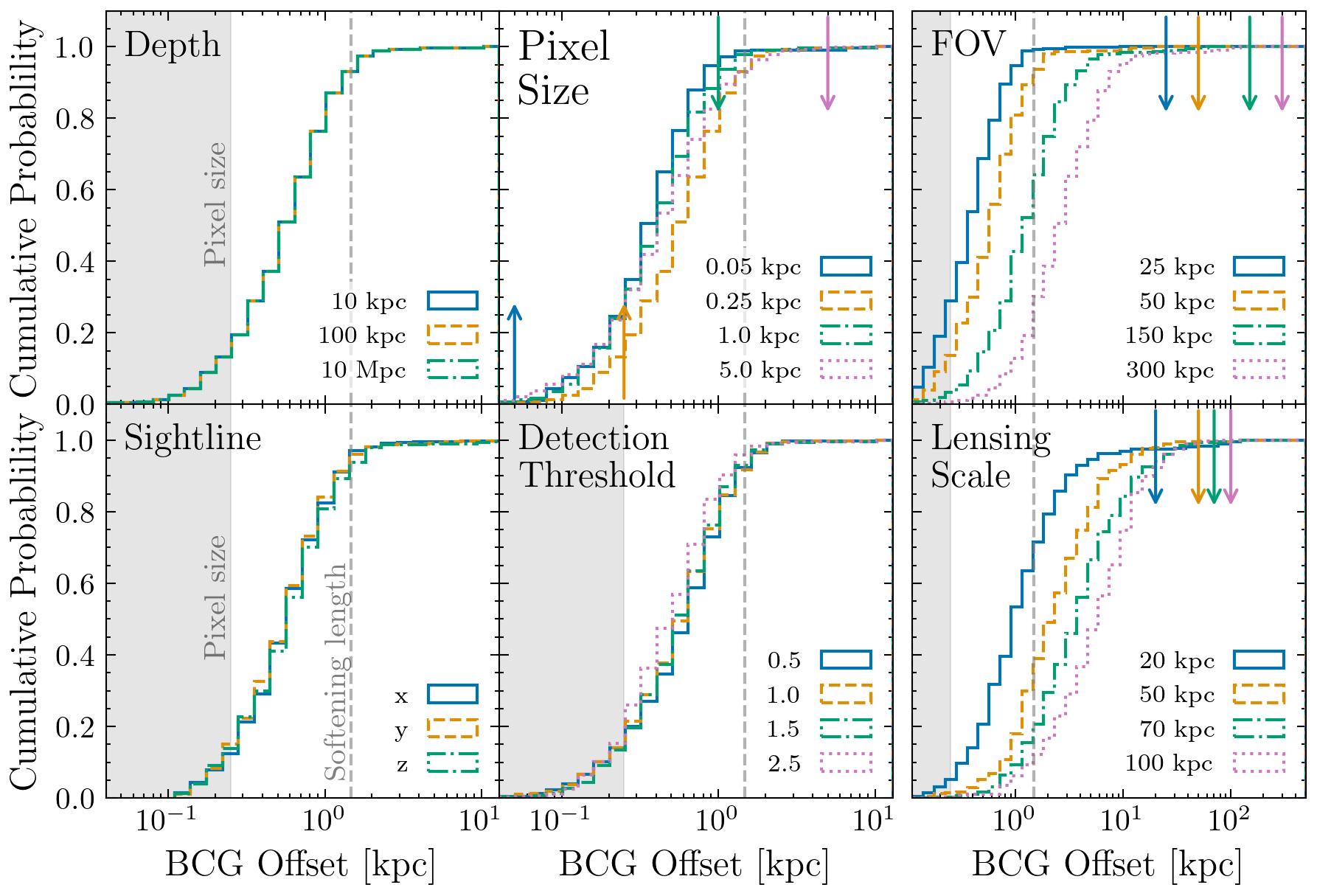}
    \caption{Nuisance parameter tests for BCG offset measurements presented as cumulative probabilities, all performed in IllustrisTNG300-1 with the fiducial parameters of Sec.~\ref{sec:massmaps} and Sec.~\ref{sec:centroid_identification} unless otherwise specified. The cumulative probability is the probability that an offset is measured as less than or equal to the corresponding $x$-axis value. All offsets calculated via SourceExtractor centroids for stars and dark matter, except for the lensing scale panel. Note the difference in scale for the rightmost column. (\emph{Upper left}) depth of the zoom box used to create the mass maps along the line of sight. (\emph{Upper middle}) pixel size used to generate mass maps. (\emph{Upper right}) side length of the zoom box used to create the mass maps. (\emph{Lower left}) sightline used for creation of mass maps, for example ``$z$'' means the mass map was calculated from the perspective of the positive $z$-axis. (\emph{Lower middle}) threshold used in SourceExtractor detection algorithm, in units of the standard deviation of the identified background. (\emph{Lower right}) Einstein radius used as a lensing scale for the area-matched isophote method. Lensing scale tests are all performed with a $300\,\rm{kpc}$ field of view. Arrows on a panel represent the value of the parameter for that offset distribution (matched in color) where appropriate.
    }
    \label{fig:hyperparameter_tests}
\end{figure*}

\subsection{Parameter Tests}\label{sec:hyperparameters}
We also examine the dependence of offsets on the various nuisance parameter choices used in this analysis, such as the parameters of the box used to create the mass maps, the parameters of source extraction thresholding procedure, or Einstein radius used in identifying the area-matched isophote. We show the results of these tests in Fig.~\ref{fig:hyperparameter_tests}, and describe the results below:

\begin{enumerate}[
  align=left,
  leftmargin=1.5em,
  itemindent=0pt,
  labelsep=0.5em,
  labelwidth=1em
]
    \item \emph{Depth}: There is some potential for intervening matter along the line of sight to increase offset measurements, but we see no evidence that this meaningfully alters the offset distribution. However, properly quantifying the effect of intervening matter requires examining the lightcone data for the group of interest, and we leave this treatment to a future study. 
    \item \emph{Pixel Size}: We test pixel sizes up to five times smaller and larger than the choice adopted in this paper ($0.25\,\rm{kpc}$ on each side), but find no dependence of the offset distribution on this choice above the softening length. Below the softening length there is a non-monotonic variance in the distribution, but this regime is unconverged and thus will not affect inference made with these results. 
    \item \emph{Field of View}: The field of view is a significant factor in the measurement of offsets via mass maps, with $\sim 50$\% of offsets above the softening scale, reaching up to $10\,\rm{kpc}$ at a field of view of $150\,\rm{kpc}$. This trend is generally described by offsets being measured up to $\lesssim 10$\% of the field of view, but this may be correlated with choices of the background identification parameters described in Sec.~\ref{sec:centroid_identification}. This correlation does not significantly alter the results of this paper, as source extraction is used in part when attempting to obtain faithful representations of the centroid for a species on small scales (where most offsets are below the softening scale regardless). It is also utilized on large scales intended to mimic real observations, for which the offsets measured via the large scale gas centroid are larger by a factor of $\sim10$ than the offsets inflated due to field of view effects in the SE-SE measurement.
    \item \emph{Sightline}: We do not expect the offset distribution to depend on the sightline used throughout all halos considered in the simulation box, and indeed we observe no evidence of a dependence.
    \item \emph{Detection Threshold}: We observe a small scatter in offset distribution as a function of the source extraction detection threshold nuisance parameter, in particular with offsets becoming slightly smaller on average as the threshold becomes more strict. This matches expectations, as this is equivalent to using only the central portions of detected objects, which likely exhibit smaller degrees of miscentering, though it should be noted that these offsets are overwhelmingly in the unconverged regime, so differences on this scale are not robustly interpretable. 
    \item \emph{Lensing Scale}: We also test the dependence of the area-matched isophote to BCG offset distribution as a function of the assumed lensing scale, all tested in the same $300\,\rm{kpc}$ field of view zoom box. Offsets increase monotonically with lensing scale, with $100\,\rm{kpc}$ lensing scale offsets on the order of $\sim 10\,\rm{kpc}$. Though this is a greatly simplified picture of lensing centers, even on scales of 20 (or 50) kpc the effect of miscentering pushes 30\% (or 60\%) of the offset distribution above the softening scale. 
\end{enumerate}

\section{Conclusions}\label{sec:conclusions}
The distribution of offsets between the BCG of a galaxy cluster and an appropriate measure of the underlying oscillation center encodes important information about the shape of the cluster core potential, and therefore the underlying dark matter particle physics. Given the challenges in identifying appropriate centers in real observations, and the unknown nature of of the dark matter underlying those observations, cosmological simulations constitute a valuable test bed for developing centroid identification methods and testing both cold dark matter and self-interacting dark matter physics. However, the usage of this technique to understand dark matter presents a number of challenges, such as the issue of appropriate centroid identification, relevant length scales being potentially unconverged and the difficulty of connecting offset distributions in simulations to sets of observed data.

We test three modern suites of cosmological hydrodynamical simulations with zero dark matter self-interaction cross section at various resolutions, namely IllustrisTNG, MillenniumTNG and BAHAMAS. We make the following observations: 
\begin{itemize}[
  align=left,
  leftmargin=\parindent,
  itemindent=0pt,
  labelsep=0.5pt,
  labelwidth=1em,
  itemsep=1em
]
    \item In each simulation, offset measurements are overwhelmingly found below the gravitational softening length when measured directly via source extraction of the smoothed particle data or using the position of the most bound particle as a measure of the cluster center, consistent with and quantifying the expectation that these cluster-scale BCG offsets should be consistent with ``zero'' in CDM. Additionally, we find no evidence for a dependence of the offset distribution on cluster mass or redshift in simulations.  
    \item When measuring centroids via the ``light traces mass'' approach, whereby gas is assumed to trace either the DM centroid or oscillation center, we find the gas to be significantly offset from the DM in cluster cores in simulations on scales above the softening length, and thus find this centroid identification technique to likely be unreliable for identifying evidence of dark matter self-interaction.
    \item We introduce a novel centroid identification technique of area-matched isophotes, intended to probe offsets on scales to which strong lensing measurements are sensitive, the method typically employed in observations. With this centroid identification method, we find offsets significantly larger than the more faithful source extraction or potential minimum offsets, increasing the median offset by a factor of $\sim 10$. This insight may serve to weaken evidence of dark matter self-interaction obtained by comparing observations to CDM simulation offset distributions. 
\end{itemize}

We also find that the baryonic physics model does contribute to some contraction of DM halos but is not clearly responsible for the overwhelmingly sub-softening offset measurements, and that the measurements are made with nuisance parameter choices in robust regimes that do not meaningfully alter the conclusions of the analysis, with the most impactful nuisance parameter (other than the scale chosen to assess miscentering relevant to lensing) being the field of view of the data considered. 

To constrain the nature of dark matter, and in particular its self-interaction cross section, it is necessary to improve the simulation data, analysis techniques and the catalogue of observations. In simulations, improvements can be made by examining high time resolution data to study the dynamics in the cores of clusters at low gravitational softening length. In the analysis of the data, it will be valuable to conduct extensive mock lensing observations in simulated clusters to establish the CDM BCG-strong lensing offset distribution. Finally, the catalogue of observations with high-quality lens models for which the BCG position is not chosen to coincide with the DM halo center has great potential to be expanded: in the near term via a homogeneous re-analysis of \emph{Hubble} archival data and highly-constrained James Webb space telescope images of strong lensing systems (eg. \cite{Cha:2024}), and in the future via large lensing surveys conducted by the Euclid and Nancy Grace Roman space telescopes.

\section*{Acknowledgments}
CR thanks Meredith Neyer for numerous helpful discussions. SB is supported by the UK Research and Innovation (UKRI) Future Leaders Fellowship [grant number MR/V023381/1].

This work makes use of the following software: \textsc{Python} \citep{python}, 
\textsc{numpy} \citep{numpy:2020}, \textsc{scipy} \citep{scipy:2020}, 
\textsc{astropy} \citep{astropy:2013, astropy:2018}, 
\textsc{jupyter} \citep{jupyter}. This research also made use of Photutils, an Astropy package for detection and photometry of astronomical sources \citep{bradley:2023:photutils}.

\bibliography{bibliography}{}
\bibliographystyle{aasjournal}

\end{document}